# Brain synchronizability, a false friend

D. Papo[1*] and J.M. Buldú[2,3]

[1]SCALab UMR CNRS 9193, Université de Lille, Villeneuve d'Ascq, France
[2]Laboratory of Biological Networks, Center for Biomedical Technology (UPM), 28223, Pozuelo de Alarcón, Madrid, Spain
[3]Complex Systems Group & G.I.S.C., Universidad Rey Juan Carlos, 28933 Móstoles, Madrid, Spain
* Email : papodav@gmail.com

**Synchronization plays a fundamental role in healthy cognitive and motor function. However, how synchronization depends on the interplay between local dynamics, coupling and topology and how prone to synchronization a network with given topological organization is are still poorly understood issues. To investigate the synchronizability of both anatomical and functional brain networks various studies resorted to the Master Stability Function (MSF) formalism, an elegant tool which allows analysing the stability of synchronous states in a dynamical system consisting of many coupled oscillators. Here, we argue that brain dynamics does not fulfil the formal criteria under which synchronizability is usually quantified and, perhaps more importantly, what this measure itself quantifies refers to a global dynamical condition that never holds in the brain (not even in the most pathological conditions), and therefore no neurophysiological conclusions should be drawn based on it. We discuss the meaning of synchronizability and its applicability to neuroscience and propose alternative ways to quantify brain networks synchronization.**

### Introduction

Consider a network in which each node is a dynamical system, e.g. an oscillator, and the links are couplings between these nodes. Can these oscillators synchronize with each other creating a coherent state and, if so, under what circumstances is this state stable? Given a particular dynamical system and coupling scheme, the *Master Stability Function* (MSF) formalism [1-3] allows relating the stability of the fully synchronized state to the spectral properties of the underlying matrix of connections, and assessing which network structures can maintain complete synchronization of the whole network.

At the macroscopic scales of typical non-invasive neuroimaging techniques, brain activity can be thought of as the collective dynamics of a set of coupled dynamical units. Synchronization among these units has been suggested to be a basic mechanism of healthy brain functioning [4]. Thus, at first glance, the problem above may seem to apply to brain activity, justifying the use of the MSF formalism to quantify brain network synchronizability. But appropriate though they may sometimes seem, formalisms are created to address very specific questions and come with their own set of formal and theoretical assumptions, the compliance with which ultimately decides whether they can be used in a given context.

In the remainder, we argue that some essential characteristics of the brain render the MSF framework difficult to apply to neuroscience, review some misunderstandings about the synchronizability construct and propose alternative ways to understand synchronization in brain networks.

### Brain synchronizability

The use of synchronizability, initially designed to study theoretical models**,** rapidly extended to the analysis of real datasets and, in the context of neuroscience, to quantify the ability of *anatomical* [5-9] and *functional* [10-23] brain networks to synchronize. For example, Tang and co-workers [9] investigated how the human brain's anatomical organization evolves from childhood to adulthood by measuring changes in the synchronizability parameter, and proposed that during the course of development human brain anatomy evolves towards an organization that limits synchronizability [9]. The authors suggested that as the brain evolves towards its mature state, it reduces its ability to synchronize, while, at the same time claim that this reduction is a consequence of promoting the controllability of the brain network. Furthermore, a few studies focused on the effects of different pathologies on brain synchronizability, such as epilepsy [14,18-23], Alzheimer's disease [8,10], or schizophrenia [17], showing statistically significant changes in the synchronizability parameter in association with these diseases. Interestingly, epilepsy was associated with an increased synchronizability during interictal activity [21], while it decreased during ictal activity [14]. Functional networks synchronizability has been reported to decrease the electroencephalographic (EEG) activity of schizophrenic patients [17]. Studies using magnetoencephalography (MEG) showed that synchronizability values depend on the frequency band considered when constructing functional networks [11].

While changes in synchronizability clearly exist, is this particular metric measuring what it is supposed to measure?

### The Master Stability Function formalism

The meaning and scope of the synchronizability construct should be understood in the MSF theoretical context it is predicated upon.

Given a group of *N* coupled dynamical systems whose dynamics in isolation follows $\dot{\mathbf{x}} = \mathbf{F}_i(\mathbf{x}_i)$, the evolution of the whole system is given by the equation:

$$\dot{\mathbf{x}}_i(t) = \mathbf{F}(\mathbf{x}_i(t)) - \sigma \sum_{j=1}^{N} l_{ij}\mathbf{H}(\mathbf{x}_j), \quad i = 1, \dots, N \quad [1]$$

where $\mathbf{x}_i$ is the n-dimensional state vector of the *ith* oscillator, $\sigma$ the coupling strength, $\mathbf{H}(\mathbf{x})$ a vectorial output function and $l_{ij}$ the elements of the Laplacian matrix *L* [24] describing how the oscillators are coupled together. For identical systems with the same coupling function $\mathbf{H}(\mathbf{x})$, the synchronized state is a solution of $\dot{\mathbf{x}}_s = \mathbf{F}(\mathbf{x}_s)$, with $\mathbf{x}_1 = \mathbf{x}_2 = \dots = \mathbf{x}_N = \mathbf{x}_s$. A linear stability analysis around the synchronization manifold allows to obtain the MSF, $\Psi(\nu)$, where the independent variable $\nu$ is related to the non-zero eigenvalues $\lambda_i$ of the Laplacian matrix as $\nu_i = \sigma\lambda_i$ [1-3]. The MSF tells how dynamics through **F** and network topology through the second term on the right side of equation [1] concur in determining the *stability* of the synchronization manifold. The term s*ynchronizability* refers to the *stability* of the *global synchronization* state.

The synchronization manifold is stable when all $\nu_i$ associated with the non-zero eigenvalues of the Laplacian matrix lie in a region in which the MSF is negative. However, different dynamical systems with different coupling functions lead to qualitatively different MSFs (see Fig. 1a for details), which can be classified as [2]: class I (always positive), class II (always negative above a threshold) and class III (negative only within a specific region). Interestingly, in the context of brain networks, synchronizability is commonly evaluated as if the brain were a class III system, although no proof of it exists. Thus, the lower the ratio $r$ between the largest and smallest (non-zero) eigenvalues (i.e. $r = \lambda_N/\lambda_2$), the more packed the eigenvalues of the Laplacian are and the highest the ability to fall within a window where the MSFs is always negative. In that sense, brain networks' synchronizability is sometimes [9] quantified by the dispersion of the eigenvalues of the Laplacian matrix $L$ [25].

*Synchronizability: some common misconceptions*

The meaning of synchronizability and the questions it allows addressing are a matter of frequent confusion and numerous misconceptions.

An important issue is whether synchronizability can be measured when ignoring the characteristics of the dynamics. Stability under perturbations exists when all eigenvalues of the combinatorial Laplacian matrix $\{\lambda_i\}$ fall within the region of stability due to the fact that the coupling is strong enough to guarantee that the MSF enters the region but weak enough to guarantee that it does not leave this region from the other side. Synchronizability is ultimately determined by the sign of the MSF evaluated at points that are indeed given by the spectrum of the Laplacian matrix and an overall coupling strength. The functional form of the MSF crucially depends on the dynamics of the coupled oscillators and the function that couples its state variables to those of other oscillators [26,27]. Depending on the shape of the MSF, dynamical systems may never synchronize, always synchronize above a certain coupling strength or synchronize only for coupling strength values within a certain range [2,27]. While the MSF for various families of dynamical systems is typically convex for generic oscillator systems, its exact shape depends not only on the dynamical systems but also on the kind of coupling between them. Thus, quantifying the synchronizability of anatomical brain networks using a parameter based on the eigenvalues of the Laplacian matrix alone, without information about the underlying dynamical oscillators and their coupling function and strength cannot ensure that the whole system falls within MSF's stability region. In other words, it is not the network structure *per se* that is synchronizable, but the particular combination of dynamical systems, coupling strength and network structure formed by the connections between these systems. While the eigenvalues of the Laplacian matrix likely contain potentially valuable information of some sort [8,10], eigenvalue dispersion of the anatomical network alone without at least some information on node dynamics cannot determine the system class one is dealing with, and conclusions on its MSF are no more than guesses (see Fig. 1a).

Two related important questions are: what are high or low synchronizability values telling us? When can synchronizability values be compared? An early study using synchronizability [11] reported that the synchronizability parameter for anatomical brain networks was close to the region where a series of theoretical models reached the synchronization manifold, based on which the authors claimed that the brain is *"located dynamically on a critical point of the order/disorder transition"* [11]. However, the bare comparisons of synchronizability values across dynamical systems and the characterization of a given topology as being more or less synchronizable than another are potentially problematic: insofar as different dynamical systems haven't necessarily got similar MSFs, the synchronizability parameter of a brain network cannot be compared with others as long as its MSF is unknown.

Perhaps at the root of most other ones, a major problematic issue relates to the frequent confusion between *synchronizability* and *synchronization*. This is for instance apparent in Tang's discussion of synchronizability [9]: "*brain networks [...] do not fully limit synchronizability, perhaps because some finite amount of synchronization is needed for dynamical coordination and cognition*" (p. 8). The synchronizability parameter does not tell if the system is synchronized or not: a system can be highly synchronizable without being synchronized, and synchronized with a low synchronizability parameter (see Fig. 1b).

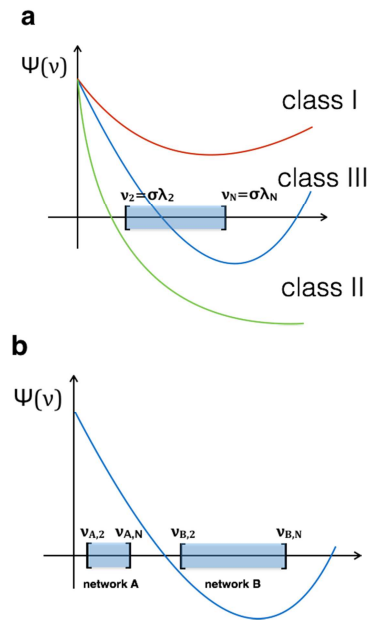

**Fig 1.** Master Stability Function $\Psi(\nu)$ as a function of the parameter $\nu$. $\nu$ is related to the (N−1) non-zero eigenvalues $\lambda_i$ of the network Laplacian matrix as $\nu_i = \sigma \lambda_i$ where $\sigma$ is the coupling strength. The synchronization manifold is stable when all $\nu_i$ lie in a region where the MSF is negative. MSFs can be classified as [2]: class I (always positive), class II (always negative above a threshold $\nu_2$) and class III (negative only within a specific region [$\nu_2, \nu_N$]). The stability region may even not be unique [27]. **(a)** Qualitative example of $\nu_i$ of a network that would synchronize class I and II dynamical systems, but not class III (assuming that $\lambda_N/\lambda_2 > \nu_N/\nu_2$). **(b)** Counter-example showing that defining synchronizability parameter as the inverse of the dispersion of the eigenvalues can be misleading: network A has lower dispersion (i.e., higher synchronizability) but lies in the region of the MSF where the synchronization manifold is unstable, while network B, has higher dispersion (and lower synchronizability) but can synchronize.

Finally, it is worth stressing that the synchronizability construct only applies to anatomical networks. This is because the MSF formalism relies on a *structural property*, i.e. the connectivity pattern between dynamical units, which should complemented by the coupling strength. However, the construction of functional networks relies on the reported coordination between brain regions, i.e. a *dynamical property*. Therefore, functional networks are not the cause of a certain level of synchronizability, but their consequence.

*Why synchronizability should not be used (in neuroscience)*

Even discounting the technical issues discussed above, fundamental reasons make the MSF-based synchronizability inapplicable to neuroscience.

Crucially, in its original formulation [1], the MSF applies to diffusively-coupled *identical* dynamical systems, i.e. all interacting units of the network should have the same variables and internal parameters. However, irrespective of the scale at which it is observed, the brain is dynamically highly heterogeneous, ruling out an application of the MSF. While the MSF formalism can be generalized to heterogeneous systems, this comes at the price of rather restrictive conditions hampering its application to brain data [26,28].

Perhaps the most fundamental obstacle to the use of the MSF in brain sciences is represented by two issues related to the definition of synchronization.

First, while various kinds of synchronization, including phase [4], generalized [29], and relay synchronization [30] have been reported for brain dynamics, and may even coexist [31], synchronizability refers to a specific synchronization mode, *complete* synchronization. Complete synchronization requires that all dynamical units have *exactly* the same phase and amplitude once the synchronization manifold is reached, a state that has never been reported in the brain (not even in its most pathological conditions).

Second, physics and neuroscience understand synchronization in fundamentally different ways: in the former, synchronization refers to a global and stable state, while in the latter to a local and transient one. While local complete synchronization may be a relevant mechanism or a reasonable modelling representation of functionally segregated regions or circuits, its dynamics is necessarily transient. Brain dynamics has in general a complex phase space geometry, and possibly no *stricto sensu* attractor at all [32,33], a scenario that cannot be dealt with using the MSF in its current form.

**Towards neurophysiologically plausible alternatives to synchronizability**

Several technical systems, e.g. power grids, wireless communication systems, require stable synchronization of their units [34-38]. Synchronizability represents, to good approximation, a construct that can be used to model and regulate their dynamics and function. However, synchronizability refers to a type of synchronization that the brain does not, certainly should not, and possibly cannot achieve in a stable way. In addition to being incompatible with the dynamical and functional heterogeneity of normal brain functioning, a completely synchronized state represents a considerable loss of complexity, and would likely be associated with an unphysiological energetic cost [39,40].

Before figuring out possible alternatives to the MSF-based synchronizability, one should perhaps address the following question: why is the MSF framework used although it so evidently at odds with neurophysiological stylized facts? What makes the MSF a convenient tool? While a unique coupling function for all network nodes and some hypotheses on the coupling matrix are convenient mathematical conditions which ensure the existence of an invariant set representing the complete synchronization manifold and considerably simplify the analysis of its stability, using steady state dynamics and complete synchronization dispenses with defining spatial topography and temporal scales of the target process.

To figure out possible alternatives to the MSF-based synchronizability construct one needs to understand both the role played by synchronization within this conceptual framework and the objective pursued by the studies using it and the problems that they aim to address. On the one hand, while in neuroscience synchronization typically refers to bivariate coupling between two neuronal ensembles, the synchronization referred to by synchronizability is in fact better thought of as a process on a network. On the other hand, from a teleological view-point, resorting to the MSF formalisms can be understood in terms of the need to address the relationship between anatomy (or, more precisely, the topology defined on it) and dynamics in complex systems [3,41,42]. Given an observed dynamics and topological organization, a construct teleologically equivalent to synchronizability may possibly be framed in terms of a networked system's propensity to enter a *functionally desirable* state or regime. But what dynamical states or regimes may represent a valuable target, the distance from which may be used as a neurophysiologically meaningful benchmark?

The true difficulty in finding alternatives to the MSF-based synchronizability is that one loses the uniqueness and task-invariance of the complete synchronization state and needs to cope with brain dynamics' spatial heterogeneity and temporal multiscaleness, and brain function's translational invariance. To define a valid equivalent of synchronizability will likely require three key ingredients: neurophysiologically plausible and functionally meaningful order parameters describing collective brain activity; mechanisms through which they may emerge; and, no less importantly, those through which they may wane. On the one hand, this should for instance involve considering networks of heterogeneous oscillators and plausible synchronization processes, compatible with metastable dynamics. On the other hand, the mechanisms through which neural assemblies interact and their role in human brain function at various scales of brain structure and dynamics should be better understood at both functional/computational and algorithmic/dynamical levels. These mechanisms are likely task-specific, and various ones may even coexist [31]. As a consequence, the definition of a dynamical target may vary as a function of the putative role of synchrony *lato sensu* in the target activity. Dynamical references could be associated with *cluster synchronization* or *chimera-like* states [43-45] which would prescribe spatial scales. However, rather than statistically stationary states, what is needed is an analysis of their dynamics, stability, bifurcations, and symmetries [44,46]. Importantly, a reference regime should also replicate the temporal scales of some (task-specific or independent) reference brain activity. The construct may for instance contain predictive information on the properties of and on the conditions under which these clusters form. Defining meaningful dynamical target processes, predicting these states and defining some sort of distance from them to given observed ones, understanding whether and the extent to which these may emerge from interactions between local dynamics and network topology are all highly non-trivial but fundamental questions, finding answers to which will likely keep the neuroscience community busy for some time to come.

**Concluding remarks**

We have argued that not only is the synchronizability construct an inadequate tool to quantify brain networks' ability to synchronize, but the problem itself to which it is supposed to provide an answer appears to be ill-posed when studying brain dynamics. More generally, the brain differs in many essential ways from the systems (e.g. the electrical power-grid or the Internet) most network theory constructs were originally designed to account for. Neuroscience, a field where network theory has only relatively recently come to the foreground [47], has so far borrowed its tools and concepts without inspiring fresh theory, and this has exposed it to the risks inherent in such an application: over-, under- and misuse of existing tools [48,49]. Rather than simply resorting to an existing bag of tricks, neuroscience should instead use the brain's unique properties to promote a fundamental reformulation of network science, for the benefit of both.

**References**

1. Pecora, L.M., & Carroll, T.L. Master stability functions for synchronized coupled systems. *Phys. Rev. Lett.* **80**, 2109–2112 (1998).
2. Boccaletti, S., Latora, V., Moreno, Y., Chavez, M., & Hwang, D.-U. Complex networks: structure and dynamics. *Phys. Rep.* **424**, 175–308 (2006).
3. Arenas, A., Díaz-Guilera, A., Kurths, J., Moreno, Y., & Zhou C.J. Synchronization in complex networks. *Phys. Rep.* **469**, 93–153 (2008).
4. Varela, F., Lachaux, J.-P., Rodriguez, E., & Martinerie, J. The brainweb: phase synchronization and large-scale integration. *Nat. Rev. Neurosci.* **2**, 229–239 (2001).
5. Chavez, M., Besserve, M., & Le Van Quyen. M. Dynamics of excitable neural networks with heterogeneous connectivity. *Progr. Biophys Mol. Bio.* **105**, 29–33 (2011).
6. Zhao, M., Zhou, C., Lü, J., & Lai, C.H. Competition between intra-community and inter-community synchronization and relevance in brain cortical networks. *Phys. Rev. E* **84**, 016109 (2011)
7. Ton, R., Deco, G., & Daffertshofer, A. Structure-function discrepancy: inhomogeneity and delays in synchronized neural networks. *PLoS Comput. Biol.* **10**, e1003736 (2014).
8. Phillips, D.J., McGlaughlin, A., Ruth, D., Jager, L.R., & Soldan, A. Graph theoretic analysis of structural connectivity across the spectrum of Alzheimer's disease: the importance of graph creation methods. *NeuroImage: Clinical* **7**, 377–390 (2015).
9. Tang, E., Giusti, C., Baum, G.L., Gu, S., Pollock, E., Kahn, A.E., Roalf, D.R., Moore, T.M., Ruparel, K., Gur, R.C., Gur, R.E., Satterthwaite, T.D., & Bassett, D.S. Developmental increases in white matter network controllability support a growing diversity of brain dynamics. *Nat. Commun.* **8**, 1252 (2017).
10. de Haan, W., van der Flier, W.M., Wang, H., Van Mieghem, P.F.A., Scheltens, P., & Stam, C.J. Disruption of functional brain networks in Alzheimer's disease: what can we learn from graph spectral analysis of resting-state magnetoencephalography? *Brain Connect.* **2**, 45–55 (2012).
11. Bassett, D.S., Meyer-Lindenberg, A., Achard, S., Duke, T., & Bullmore, E. Adaptive reconfiguration of fractal small-world human brain functional networks. *Proc. Natl. Acad. Sci. U.S.A.* **51**, 19518–19523 (2006).
12. Reijneveld, J.C., Ponten, S.C., Berendse, H.W., & Stam, C.J. (2007). The application of graph theoretical analysis to complex networks in the brain. *Clin. Neurophysiol.* **118**, 2317–2331.
13. Stam, C.J., & Reijneveld, J.C. Graph theoretical analysis of complex networks in the brain. *Nonlin. Biomed. Phys.* **1**, 3 (2007).
14. Schindler, K.A., Bialonski, S., Horstmann, M.T., Elger, C.E., & Lehnertz, K. Evolving functional network properties and synchronizability during human epileptic seizures. *Chaos* **18**, 033119 (2008).
15. Deuker, L., Bullmore, E.T., Smith, M., Christensen, S., Nathan, P.J., Rockstroh, B., & Bassett, D.S. Reproducibility of graph metrics of human brain functional networks. *NeuroImage* **47**, 1460–1468 (2009).
16. van Wijk, B.C.M., Stam C.J., & Daffertshofer, A. Comparing brain networks of different size and connectivity density using graph theory. *PLoS ONE* **5**, e13701 (2010). doi:10.1371/journal.pone.0013701
17. Jalili, M., & Knyazeva, M.G. EEG-based functional networks in schizophrenia. *Comput. Biol. Med.* **41**, 1178–1186 (2011).
18. van Dellen, E., Douw, L., Hillebrand, A., Ris-Hilgersom, I.H.M., & Schoonheim, M.M., et al. MEG network differences between low- and high-grade glioma related to epilepsy and cognition. *PLoS ONE* **7**, e50122 (2012).
19. Tahaei, M.S., Jalili, M., & Knyazeva, M.G. Epilepsy synchronizability of EEG-based functional networks in early Alzheimer's disease. *IEEE Trans. Neural Syst. Rehabil. Eng.* **5**, 636–641 (2012).
20. Bialonski, S., Lehnertz, K. Assortative mixing in functional brain networks during epileptic seizures. *Chaos* **3**, 033139 (2013) doi: 10.1063/1.4821915.
21. Lehnertz, K., Ansmann, G., Bialonski, S., Dickten, H., Geier, C., & Porz, S. Evolving networks in the human epileptic brain. *Physica D* **267**, 7–15 (2014).
22. Niso, G., Carrasco, S., Gudín, M., Maestú, F., del-Pozo, F., & Pereda, E. What graph theory actually tells us about resting state interictal MEG epileptic activity. *NeuroImage*: *Clinical* **8**, 503–515 (2015).
23. Khambhati, A.N., Davis, K. A., Lucas, T.H., Litt, B., & Bassett, D.S. Virtual cortical resection reveals push-pull network control preceding seizure evolution. *Neuron* **91**, 1170–1182 (2016).
24. The (combinatorial) Laplacian matrix is defined as $L = D - A$, where $D$ is a diagonal matrix containing the degree (i.e., number of connections) of the nodes of the network and $A$ is the adjacency matrix, with $A_{ij} = 1$ if nodes $i$ and $j$ are connected and zero otherwise. An adaptation to weighted networks can easily be obtained by including the weights of the connections in $A$ and replacing the node degree by the node strength (i.e., the sum of the weights of node's links).
25. The dispersion of the eigenvalues of the Laplacian matrix is given the following expression: $1/s^2 = \sigma^2 (N-1)/\sum_{i=1}^{N-1} |\lambda_i - \bar{\lambda}|^2$ with $\bar{\lambda} = (1/N - 1)\sum_{i=1}^{N-1} \lambda_i$, $\sigma$ being the coupling strength of the $N$ nodes in the anatomical network and $\lambda_i$ the i-th eigenvalue of the Laplacian matrix.
26. Nishikawa, T., & Motter, A.E. Network synchronization landscape reveals compensatory structures, quantization, and the positive effect of negative interactions. *Proc. Natl. Acad. Sci. USA* **107**, 10342–10347 (2010).
27. Huang, L., Chen, Q., Lai, Y.-C., & Pecora, L.M. Generic behavior of master-stability functions in coupled nonlinear dynamical systems. *Phys. Rev. E* **80**, 036204 (2009).
28. Sun, J., Bollt, E.M., & Nishikawa, T. Master stability functions for coupled nearly identical dynamical systems. *EPL* **85**, 60011 (2009).
29. Stam, C.J., & van Dijk, B.W. Synchronization likelihood: an unbiased measure of generalized synchronization in multivariate data sets. *Physica D* **163**, 236–251 (2002).
30. Vicente, R., Gollo, L.L., Mirasso, C.R., Fischer, I., & Pipa, G. Dynamical relaying can yield zero time lag neuronal synchrony despite long conduction delays. *Proc. Natl. Acad. Sci. USA* **105**, 17157–17163 (2008).
31. Malagarriga, D., Villa, A.E., García-Ojalvo, J., & Pons, A.J. Consistency of heterogeneous synchronization patterns in complex weighted networks. *Chaos* **27**, 031102 (2017).
32. Rabinovich, M., Huerta, R., & Laurent, G. Transient dynamics for neural processing. *Science* **321**, 48–50 (2008).
33. Tognoli, E., & Kelso, J.A.S. The metastable brain. *Neuron* **81**, 35–48 (2014).
34. Kinzel, W., Englert, A., & Kanter, I. On chaos synchronization and secure communication. *Phil. Trans. R. Soc. A* **368**, 379–389 (2010).
35. Tyrrell, A., Auer, G., & Bettstetter, C. Emergent slot synchronization in wireless networks. *IEEE Trans. Mobile Comput.* **9**, 719–732 (2010).
36. Rohden, M., Sorge, A., Timme, M., & Witthaut, D. Self-organized synchronization in decentralized power grids. *Phys. Rev. Lett.* **109**, 064101 (2012).
37. Rohden, M., Sorge, A., Witthaut, D., & Timme, M. Impact of network topology on synchrony of oscillatory power grids. *Chaos* **24**, 013123 (2014).
38. Motter, A.E., Myers, S.A., Anghel, M. & Nishikawa, T. Spontaneous synchrony in power-grid networks. *Nat. Phys.* **9**, 191–197 (2013).
39. Torrealdea, F.J., Sarasola, C., d'Anjou, A., Moujahid, A., & de Mendizábal, N.V. Energy efficiency of information transmission by electrically coupled neurons. *BioSystems* **97**, 60–71 (2009).
40. Moujahid, A., d'Anjou, A., Torrealdea, F.J., & Torrealdea, F. Energy and information in Hodgkin-Huxley neurons. *Phys. Rev. E* **83**, 031912 (2011).
41. Skardal, P.S., Taylor, D., & Sun, J. Optimal synchronization of complex networks. *Phys. Rev. Lett.* **113**, 144101 (2014).
42. Menck, P.J., Heitzig, J., Marwan, N., & Kurths, J. How basin stability complements the linear-stability paradigm. *Nat. Phys.* **9**, 89–92 (2013).
43. Zhou, C., & Kurths, J. Hierarchical synchronization in complex networks with heterogeneous degrees. *Chaos* **16**, 015104 (2006).
44. Abrams, D.M., Mirollo, R., Strogatz, S.H., & Wiley, D.A. Solvable model for chimera states of coupled oscillators. *Phys. Rev. Lett.* **101**, 084103 (2008).
45. Bi, H., Hu, X., Boccaletti, S., Wang, X., Zou, Y., Liu, Z., & Guan, S. Coexistence of quantized, time dependent, clusters in globally coupled oscillators. *Phys. Rev. Lett.* **117**, 204101 (2016).
46. Pecora, L.M., Sorrentino, F., Hagerstrom, A.M., Murphy, T.E., & Roy, R. Cluster synchronization and isolated desynchronization in complex networks with symmetries. *Nat. Commun.* **5**, 4079 (2014).
47. Bullmore, E., & Sporns, O. Complex brain networks: Graph theoretical analysis of structural and functional systems, *Nat. Rev. Neurosci.* **10**, 186–198 (2009).
48. Papo, D., Zanin, M., Pineda-Pardo, J.A., Boccaletti, S., & Buldú, J.M. Functional brain networks: great expectations, hard times, and the big leap forward. *Phil. Trans. R. Soc. B* **369**, 20130525 (2014).
49. Papo, D., Zanin, M., Martínez, J.H., & Buldú, J.M. Beware of the Small-World neuroscientist! *Front. Hum. Neurosci.* **10**, 96 (2016).